\documentclass[letter]{aa}                                          
\usepackage{array}                                                  
\usepackage{txfonts,epsfig,graphicx,natbib,url,twoopt}
\usepackage[breaklinks=true]{hyperref}                              
\usepackage[labelfont=footnotesize,textfont=footnotesize]{caption}  
\bibpunct{(}{)}{;}{a}{}{,}    
\newcommandtwoopt{\citeads}[3][][]{\href{http://adsabs.harvard.edu/abs/#3}%
                                        {\citealp[#1][#2]{#3}}}
\newcommandtwoopt{\citepads}[3][][]{\href{http://adsabs.harvard.edu/abs/#3}%
                                         {\citep[#1][#2]{#3}}}
\newcommandtwoopt{\citetads}[3][][]{\href{http://adsabs.harvard.edu/abs/#3}%
                                         {\citet[#1][#2]{#3}}}
\newcommandtwoopt{\citeyearads}
  [3][][]{\href{http://adsabs.harvard.edu/abs/#3}%
  {\citeyear[#1][#2]{#3}}}
\begin{document}
   \title{
          INTEGRAL observations of Sco{\thinspace}X-1: evidence \\
          for Comptonization up to $200${\thinspace}keV
         }

   \subtitle{}

   \author{
          T. Maiolino
          \inst{ }
          \and
          F. D'Amico
          \inst{ }
          \and
          J. Braga
          \inst{ }
          }

   \institute{
              Instituto Nacional de Pesquisas Espaciais -- INPE
              Avenida dos Astronautas 1758, \\
              12227-010, S\~ao Jos\'e dos Campos-SP, Brazil \\
              \label{1}
              \email{tais.maiolino@das.inpe.br;damico@das.inpe.br;braga@das.inpe.br}
             }

   \date{ }

\abstract{
            We have analyzed a long-term database for Sco{\thinspace}X-1 obtained with
            the telescope IBIS onboard the INTEGRAL satellite in order
            to study the hard X-ray behavior of Sco{\thinspace}X-1 from $20$ up to
            $200${\thinspace}keV. Besides the data used for producing of the
            INTEGRAL catalog of sources, this is the longest ($412${\thinspace}ks)
            database of IBIS on Sco{\thinspace}X-1 up to date. The production of hard
            X-ray tails in low-mass X-ray binaries is still a matter of debate.
            Since most of the fits to the high-energy part of the
            spectra are done with powerlaw models, the physical mechanism for the
            hard X-ray tail production is unclear. The purpose of this study is
            to better constrain those possible mechanisms. Our main result shows a
            strong correlation between the fluxes in the thermal and nonthermal
            part of Sco{\thinspace}X-1 spectra. We thus suggest that Comptonization of
            lower energy photons is the mechanism for producing hard X-ray
            tails in Sco{\thinspace}X-1.
         }
			 
               
\maketitle
%

\section{Introduction}
\label{intro}

INTEGRAL (\citeads{2003A&A...411L...1W}) has played a key role in the
observations of cosmic X and ${\gamma}$-ray sources over the past
decade.  Our study here is a long time-span ($\sim${\thinspace}7 years)
observation of Sco{\thinspace}X-1 (\citeads{1962PhRvL...9..439G}). INTEGRAL has the
maximum sensitivity for observing Sco{\thinspace}X-1 in energies above
$\sim${\thinspace}30{\thinspace}keV. Our database was chosen to maximize the
sensitivity in our timespan, and it is the longest sample of
IBIS/INTEGRAL observations of Sco{\thinspace}X-1, besides the $4$th
INTEGRAL catalog (\citeads{2010ApJS..186....1B}).  Such a homogeneous database 
has been important for reducing any systematic error due to
(e.g.) different flux normalization factors when comparing data from
two different instruments.

Sco{\thinspace}X-1 is a {\sf Z} source. The {\sf Z} sources are low-mass X-ray binaries
(LMXBs) fed by accretion onto a neutron star with a low magnetic field
(see, e.g, \citeads{2006csxs.book...39V}).  In these systems an accretion disk is
always present, making the {\sf Z}s persistent sources at the hard
X-ray domain ($E \; {{\lower.5ex\hbox{$\; \buildrel > \over \sim $}}} \;20${\thinspace}keV) 
up to $40${\thinspace}keV.  The
spectra of the {\tt Z}s can be divided into two distinct components: a 
{\it  thermal} one, which dominates the emission up to
$\sim$40--50{\thinspace}keV, and a {\it nonthermal}, highly variable
component, which dominates the spectrum at higher energies.  The
origin of this highly variable component above $\sim${\thinspace}50 keV is still a
matter of debate. Hard X-ray spectra of {\sf Z} sources have already been
extensively reported 
(e.g., \citeads{2000ApJ...544L.119D};  \citeads{2001ApJ...554...49D};
\citeads{2001ApJ...547L.147D}; \citeads{2001AdSpR..28..389D}; \citeads{2004ApJ...600..358I}; 
\citeads{2004NuPhS.132..616L}). 
Sco{\thinspace}X-1 plays a key role in  understanding these sources, and it was
observed well with INTEGRAL (e.g., \citeads{2006ApJ...649L..91D}; 
\citeads{2006A&A...459..187P}).

In this work, based in a long-term sample of Sco{\thinspace}X-1
INTEGRAL observations, 
we show evidence of the Comptonization of (softer) photons up to
energies $\sim${\thinspace}200{\thinspace}keV. This is one of the few proofs 
to date of Comptonization at $E {{\lower.5ex\hbox{$\; \buildrel > \over \sim $}}} \;
50${\thinspace}keV for Sco{\thinspace}X-1.


\section{Data selection and analysis}
\label{data}

The data presented here were collected by the IBIS (Imager on Board the Integral Satellite)
telescope (\citeads{2003A&A...411L.131U}) 
onboard INTEGRAL. It is noteworthy that the JEM-X
(Joint European Monitor for X-rays, \citeads{2003A&A...411L.231L}) 
count rate is saturated (in almost every practical
situation) by Sco{\thinspace}X-1. In this way, not having coverage in the
low-energy X-ray band of JEM-X, we lost the ability to track
Sco{\thinspace}X-1 movement along the {\tt Z} diagram. Such a movement might be
driven by the mass accretion rate (e.g. \citeads{1996ApJ...469L...1V}). Following
the recommendations of the IBIS cookbook, we adopted a 2{\%} systematic error in
the IBIS data.  This work made use of the version 9.0 of the
INTEGRAL/OSA software (e.g., \citeads{2003A&A...411L.223G}). 
We took into account, as stressed by IBIS cookbook, the need to extract
the spectra of all the sources in the field of view (FOV)
simultaneously to avoid possible any sources of errors. Only sources with a
signal-to-noise ratio (S/N) greater than
$\sim${\thinspace}6 in the FOV needed to be included for the (time-consuming)
spectrum extraction.  Taking a conservative approach and considering 
to the relative emptiness of the FOV around Sco{\thinspace}X-1, we included all
the sources with S/N{\thinspace}$>${\thinspace}2 for the extraction process.

Data were selected based on two primary goals: (1) select the longest
observations to maximize IBIS sensitivity; and (2) select
(public) observations from 2003 to 2010, creating a long-term
database. Our spectral analysis in XSPEC clearly indicates the need
for a two-component model. For the first component (referred to here as the
{\it thermal} component), we used the {\tt comptt} model in XSPEC. It may
sound unusual to use a model that is based on the Comptonization of
(seed) blackbody Wien's photons (\citeads{1994ApJ...434..570T}) and to call it 
{\it  thermal}. We just want to emphasize that this part of the Sco{\thinspace}X-1
spectra can be adjusted by thermal models {\it and} that the 
{\tt  comptt} uses a thermal seed for photons. 

Actually, fits with
a simple blackbody also provide adequate fits to the thermal part of our
spectra (20--50{\thinspace} keV). In our database, nevertheless, there is a
subset that was analyzed (\citeads{2006ApJ...649L..91D}) using the 
{\tt  comptt} model. In such a way, to maintain uniformity within our
study and to make comparisons with other observations,
we decided to use the {\tt comptt} model. This
thermal component of the spectrum extends to energies up to
$\sim${\thinspace}40--50{\thinspace}keV.  The need for including of a second
component in the spectrum can be easily verified by (1) the residuals
of the fit using only the {\it thermal} part and by (2) the significant
S/N of the channels from $\sim${\thinspace}50--200{\thinspace}keV. 

For this second component, a simple {\tt pegpw} model in XSPEC adjusts
the spectra very well .  It is interesting, however, to note that this
second part of the spectrum can also be fitted as a {\it{second}}
Comptonization in the spectrum. Accordingly, we performed the fits for
the spectra in our database with two models: (1) {\tt{comptt + pegpw}}
and (2) {\tt{comptt+comptt}}, both of them providing adequate
fits. We noticed that, in general the fits
with {\tt{comptt+pegpw}} are better, so we interpreted our results
here, taking this model into account. 

\section{Results}
\label{results}

\begin{figure}
\resizebox{\hsize}{!}{\includegraphics[angle=270]{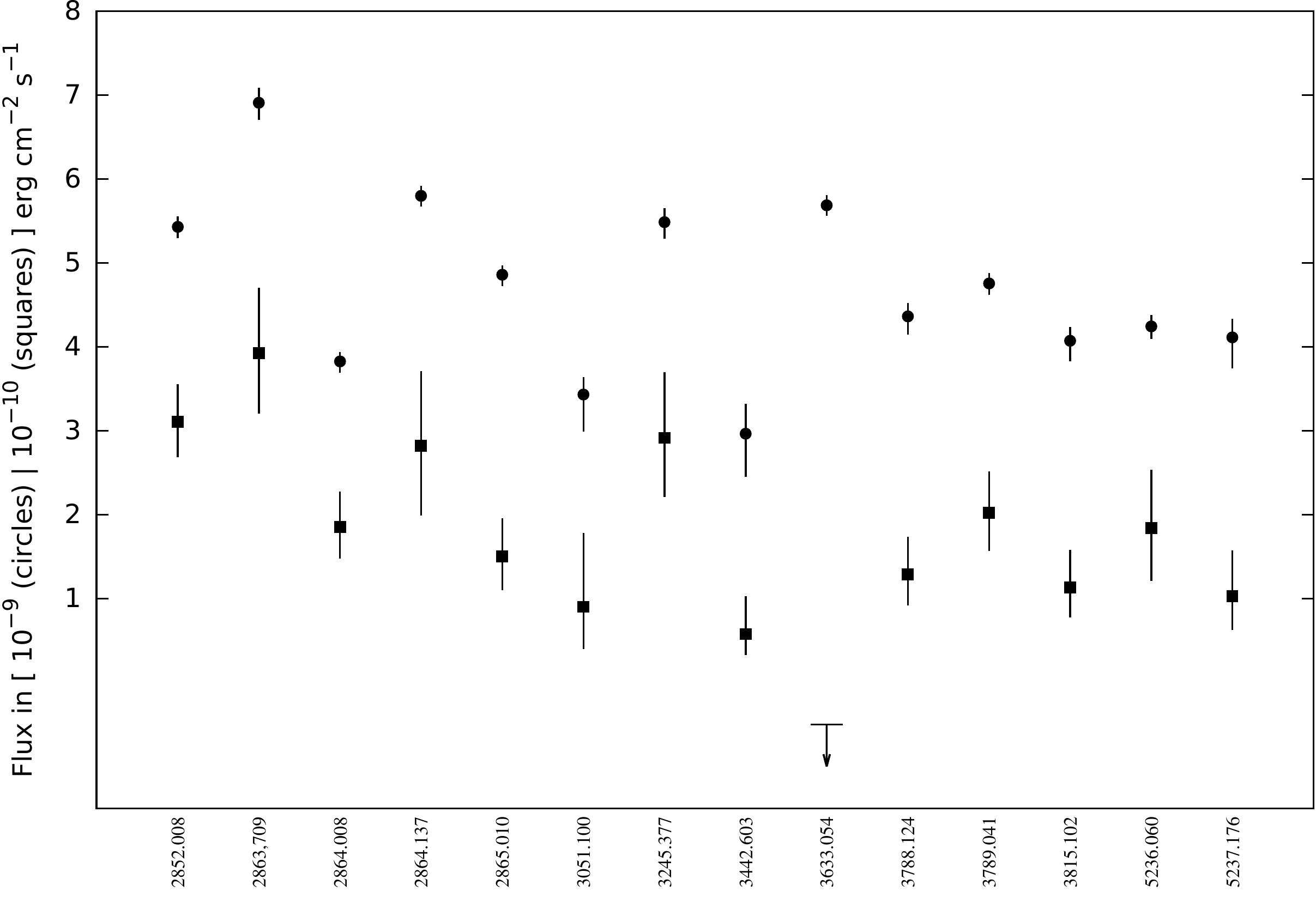}}
		\caption{
                         The INTEGRAL/ISGRI thermal (20-50{\thinspace}keV: circles) and powerlaw 
                         (20-200{\thinspace}keV: squares)  fluxes for Sco{\thinspace}X-1 plotted 
                         against MJD{\thinspace}$-${\thinspace}50000 (as derived from OSA software). 
                         Data span of 7 years of observations (from 2003 to 2010). The arrow is 
                         a 3{\thinspace}$\sigma$ upper limit of $\sim \; 5 \; \times \; 10^{-12}$ in 
                         the 50 to 200{\thinspace}keV flux.
        	        }
\label{fig1}
\end{figure}

Table ({\ref{tab1}}) summarizes key results of our data analysis.
In Figure ({\ref{fig1}}) a time history of the two fluxes (thermal
and powerlaw) is shown for our database. Only on one occasion
(MJD=53633.054) was the powerlaw component not necessary to
achieve a good fit. This absence of the powerlaw
component (only 1 out of 14 observations) for Sco{\thinspace}X-1 represents a
variability to a lesser degree than observed with RXTE/HEXTE 
(High Energy X-ray Timing Experiment) long-term
observations (\citeads{2001ApJ...547L.147D}; \citeads{2001AdSpR..28..389D}).  It is
unclear what causes the powerlaw component to disappear.

\begin{table*}
\caption{INTEGRAL/IBIS/ISGRI observations of Sco{\thinspace}X-1}
\label{tab1}
\begin{tabular}{cccccccccc}
\hline\noalign{\smallskip}
                      &                              &     & \multicolumn{2}{c}{\textsc{compTT}}    &  & \multicolumn{2}{c}{\textsc{PEGPW}}    &  &                    \\ \cline{4-5} \cline{7-8} 
Start of observation  & Livetime \tablefootmark{(a)} & MJD & KT$_e$ (keV) & Flux\tablefootmark{(b)} &  & ${\Gamma}$  & Flux\tablefootmark{(c)} & Flux\tablefootmark{(d)} & ${\chi}^2_{\hbox{red}}$\tablefootmark{(e)} \\ \hline
\vspace{2mm}
$31/07/2003$          & 4.92 &52852.008&$2.86^{+0.04}_{-0.04}$ & $5.43^{+0.13}_{-0.14}$ & & $2.62^{+0.29}_{-0.26}$ & $3.11^{+0.45}_{-0.43}$         & $3.01^{+0.43}_{-0.42}$ &$0.6/0.8$               \\
\vspace{2mm} 
$11/08/2003$          & 1.47 &52863.709&$3.06^{+0.05}_{-0.05}$ & $6.91^{+0.18}_{-0.20}$ & & $2.63^{+0.38}_{-0.34}$ & $3.92^{+0.78}_{-0.72}$         & $3.79^{+0.84}_{-0.80}$ &$0.8/0.6$               \\
\vspace{2mm}
$12.008/08/2003$      & 4.94 &52864.008&$2.81^{+0.05}_{-0.05}$ & $3.38^{+0.11}_{-0.13}$ & & $2.90^{+0.44}_{-0.38}$ & $1.86^{+0.42}_{-0.38}$         & $1.75^{+0.39}_{-0.37}$&$0.5/0.7$               \\
\vspace{2mm}
$12.137/08/2003$      & 1.55 &52864.137&$2.85^{+0.04}_{-0.04}$ & $5.80^{+0.12}_{-0.13}$ & & $1.87^{+0.58}_{-0.48}$ & $2.82^{+0.89}_{-0.83}$         & $2.93^{+1.22}_{-0.88}$&$0.6/0.6$               \\
\vspace{2mm}
$13/08/2003$          & 4.38 &52865.010&$2.80^{+0.04}_{-0.04}$ & $4.86^{+0.11}_{-0.13}$ & & $2.69^{+0.59}_{-0.48}$ & $1.50^{+0.45}_{-0.40}$         & $1.46^{+0.46}_{-0.21}$&$0.4/0.4$               \\
\vspace{2mm}
$15/02/2004$          & 9.42 &53051.100&$2.66^{+0.08}_{-0.01}$ & $3.43^{+0.21}_{-0.44}$ & & $3.60^{+1.56}_{-1.21}$ & $0.90^{+0.88}_{-0.51}$         & $0.78^{+0.91}_{-0.52}$&$0.5/0.5$               \\
\vspace{2mm}  
$27/08/2004$          & 1.64 &53245.377&$2.71^{+0.06}_{-0.07}$ & $5.49^{+0.17}_{-0.20}$ & & $2.78^{+0.50}_{-0.42}$ & $2.92^{+0.78}_{-0.70}$         & $2.75^{+0.77}_{-0.67}$&$1.4/1.7$                \\
\vspace{2mm}
$12/03/2005$          & 1.22 &53442.603&$2.60^{+0.08}_{-0.09}$ & $2.96^{+0.36}_{-0.52}$ & & $4.75^{+1.06}_{-1.17}$ & $0.58^{+0.45}_{-0.25}$         & $0.50^{+0.43}_{-0.23}$&$1.0/0.9$                \\
\vspace{2mm}
$19/09/2005$          & 1.21 &53633.054&$2.36^{+0.27}_{-0.03}$ & $5.69^{+0.12}_{-0.12}$ & &                     &$<0.05$ \tablefootmark{(f)}    & &$1.4$\tablefootmark{(g)}                  \\
\vspace{2mm}
$21/02/2006$          & 3.80 &53788.124&$2.61^{+0.06}_{-0.07}$ & $4.36^{+0.16}_{-0.22}$ & & $3.41^{+0.68}_{-0.56}$ & $1.29^{+0.44}_{-0.37}$         & $1.23^{+0.43}_{-0.41}$&$1.1/1.2$                 \\
\vspace{2mm}
$22/02/2006$          & 4.21 &53789.041&$2.48^{+0.07}_{-0.09}$ & $4.75^{+0.13}_{-0.14}$ & & $2.83^{+0.43}_{-0.37}$ & $2.02^{+0.50}_{-0.45}$         & $1.92^{+0.50}_{-0.45}$&$1.6/1.8$                 \\
\vspace{2mm}
$20/03/2006$          & 3.76 &53815.102&$2.65^{+0.05}_{-0.06}$ & $4.07^{+0.17}_{-0.24}$ & & $3.51^{+0.77}_{-0.64}$ & $1.13^{+0.44}_{-0.36}$         & $1.00^{+0.46}_{-0.35}$&$1.0/1.1$                 \\
\vspace{2mm}
$08/02/2010$          & 3.53 &55236.060&$2.73^{+0.07}_{-0.08}$ & $4.24^{+0.13}_{-0.15}$ & & $2.20^{+0.72}_{-0.58}$ & $1.84^{+0.69}_{-0.63}$         & $1.79^{+0.72}_{-0.65}$&$1.6/1.6$                 \\
\vspace{2mm}
$09/02/2010$          & 3.60 &55237.176&$2.74^{+0.08}_{-0.09}$ & $4.11^{+0.22}_{-0.37}$ & & $3.55^{+1.07}_{-0.86}$ & $1.03^{+0.55}_{-0.40}$         & $0.89^{+0.57}_{-0.42}$&$1.0/0.9$                 \\
\noalign{\smallskip}\hline
\end{tabular}
\tablefoot{
           Uncertainties at 90{\%} confidence level for the derived parameters of the model applied.\\
           \tablefoottext{a}{IBIS livetime in units of 10$^4$s.}
           \tablefoottext{b}{
                             Flux of the thermal component in the 20--50 keV band in units of 
                             10$^{-9}${\thinspace}erg{\thinspace}cm$^{-2}${\thinspace}s$^{-1}$.
                            }
           \tablefoottext{c}{
                             Flux of the nonthermal component the 50--200 keV band in units of 
                             10$^{-10}${\thinspace}erg{\thinspace}cm$^{-2}${\thinspace}s$^{-1}$.
                            } 
           \tablefoottext{d}{
                             Flux of a comptt model applied at the 50--200 keV band in units of
                             10$^{-10}${\thinspace}erg{\thinspace}cm$^{-2}${\thinspace}s$^{-1}$.
                            }
           \tablefoottext{e}{model=comptt+pegpw/model=comptt+comptt (see text).}
           \tablefoottext{f}{3${\sigma}$ upper limit on the (pegpw) flux.}
           \tablefoottext{g}{model=comptt alone: 50--200 keV component is absent (see text).}
          }
\end{table*}

Observations from the year 2000 up to now of Sco{\thinspace}X-1 are shown
in Table ({\ref{tab2}}). It can be seen that our database presented here is 
the longest one for Sco using INTEGRAL/IBIS, besides the fourth INTEGRAL catalog
itself.

\begin{table*}
\setlength{\extrarowheight}{1ex}
\caption{Modern observations of Sco{\thinspace}X-1}
\label{tab2}
\begin{tabular}{c c c c c c c c}
\hline\noalign{\smallskip}
                          & \multicolumn{2}{c}{[HEXTE/RXTE]} & [SPI/INTEGRAL] &                 \multicolumn{4}{c}{[IBIS/INTEGRAL]}
                            \\ \cline{2-3} \cline{5-8}
     & 2001a\tablefootmark{(a)}   & 2001b\tablefootmark{(b)}     & 2003\tablefootmark{(c)}    & 2006a\tablefootmark{(d)}    & 
       2006b\tablefootmark{(e)}   & 2010\tablefootmark{(f)}      & This study       \\ \hline
Exposure Time\tablefootmark{(g)}                &   1.04         &   2.03       &  9.58         &   2.66      &   3.72     &  63.0 & 4.12                 \\
F$_{\hbox{\scriptsize{Th}}}^{\hbox{\scriptsize{20-50 keV}}}$\tablefootmark{(h)} &  7.82$\pm$0.68 & 7.82$\pm$62  &           &            &       &      & 4.71$\pm$0.28 \\
F$_{\hbox{\scriptsize{total}}}^{\hbox{\scriptsize{20-40 keV)}}}$\tablefootmark{(i)} &   &              & 3.47$\pm$0.03 & 5.8\tablefootmark{(m)} & 6.1$\pm$0.2  & 4.46$\pm$0.15  & 
                                                                                                                                                    4.69$\pm$0.28 \\
F$_{\hbox{\scriptsize{total}}}^{\hbox{\scriptsize{40-200 keV}}}$\tablefootmark{(j)} &             &              &           &            & 2.6$\pm$1.8   &    & 2.74$\pm$0.34 \\
F$_{\hbox{\scriptsize{total}}}^{\hbox{\scriptsize{40-150 keV}}}$\tablefootmark{(k)} &             &              &           & 2.2\tablefootmark{(m)} &  &     & 2.48$\pm$0.29 \\
F$_{\hbox{\scriptsize{PL}}}^{\hbox{\scriptsize{20-200 keV}}}$\tablefootmark{(l)}   & 10.7$\pm$1.6 & 10.0$\pm$1.2 &          &            & 6.27$\pm$0.35 &     & 5.57$\pm$0.47 \\
averaged value of ${\Gamma}$         & 0.98$\pm$0.60  & 0.84$\pm$0.41&               &3.85$^{+0.01}_{-0.08}$&3.31$^{+0.08}_{-0.17}$     &        & 3.03$\pm$0.20 \\
\noalign{\smallskip}\hline
\end{tabular}
\tablefoot{
\tablefoottext{a}{\citeads{2001ApJ...547L.147D}}
\tablefoottext{b}{\citeads{2001AdSpR..28..389D}}
\tablefoottext{c}{\citeads{2003A&A...411L.363P}}
\tablefoottext{d}{\citeads{2006A&A...459..187P}}
\tablefoottext{e}{\citeads{2006ApJ...649L..91D}}
\tablefoottext{f}{\citeads{2010ApJS..186....1B}}
\tablefoottext{g}{in 10$^5$ s.}
\tablefoottext{h}{flux of the thermal component, 20--50 keV, in units of 10$^{-9}${\thinspace}erg{\thinspace}cm$^{-2}${\thinspace}s$^{-1}$.}
\tablefoottext{i}{20 to 40 keV flux, in units of 10$^{-9}${\thinspace}erg{\thinspace}cm$^{-2}${\thinspace}s$^{-1}$.}
\tablefoottext{j}{40 to 200 keV flux, in units of 10$^{-10}${\thinspace}erg{\thinspace}cm$^{-2}${\thinspace}s$^{-1}$.}
\tablefoottext{k}{40 to 150 keV flux, in units of 10$^{-10}${\thinspace}erg{\thinspace}cm$^{-2}${\thinspace}s$^{-1}$.}
\tablefoottext{l}{flux of the powerlaw component, 20 to 200 keV, in units of 10$^{-9}${\thinspace}erg{\thinspace}cm$^{-2}${\thinspace}s$^{-1}$.}
\tablefoottext{m}{quoted with no errors in the original reference.}
}
\end{table*}

From the results shown in Table{\thinspace}({\ref{tab1}}), the average fluxes are
F$_{\hbox{\tiny{20-50 keV}}}^{\hbox{\tiny{thermal}}} \, = \, 4.87 \; \pm 0.81 \; \times \; 10^{-9}${\thinspace}erg{\thinspace}cm$^{-2}${\thinspace}s$^{-1}$
and
F$_{\hbox{\tiny{50-200 keV}}}^{\hbox{\tiny{powerlaw}}} \, = \, 1.76 \; \pm 0.84 \; \times \; 10^{-10}${\thinspace}erg{\thinspace}cm$^{-2}${\thinspace}s$^{-1}$.
Using the 2.8{\thinspace}kpc distance to Sco{\thinspace}X-1 
(\citeads{2001ApJ...558..283F}), we have
L$_{\hbox{\tiny{thermal}}} \; = \; 4.6 \; \pm 0.8 \; \times \; 10^{36}${\thinspace}erg{\thinspace}s$^{-1}$ 
and
L$_{\hbox{\tiny{powerlaw}}} \; = \; 1.6 \; \pm 0.8 \; \times \; 10^{35}${\thinspace}erg{\thinspace}s$^{-1}$.
For comparison, the thermal flux given by 
\citetads{2001AdSpR..28..389D} was
F$_{\hbox{\tiny{20-50 keV}}}^{\hbox{\tiny{th,HEXTE}}} \, = \, 7.1 \; \pm 1.8 \; \times \; 10^{-9}${\thinspace}erg{\thinspace}cm$^{-2}${\thinspace}s$^{-1}$. 
characterizing a high degree of variability in the thermal component.

In Table ({\ref{tab1}}) we also list the fluxes in the 50--200{\thinspace}keV derived from the
{\tt{comptt + comptt}} model. For our database the average plasma temperature of the second {\tt{comptt}}
was found to be $60.4${\thinspace}$\pm${\thinspace}$1.1$ keV, as would be expected to extend the fit
up to $\sim${\thinspace}$200${\thinspace}keV (\citeads{2008ApJ...680..602F}).
It is also important to highlight that the temperature
of the seed (Wien) photons for this (second) Comptonization are on order of 1{\thinspace}keV,
again as we would expect.

Although variability can be noticed from the results displayed in
Table ({\ref{tab2}}), it is noteworthy that, taking
only INTEGRAL observations into account (columns 4--8 in the Table{\thinspace}{\ref{tab2}})
the degree of variability is not extreme (exception to this is the total
20--40{\thinspace}keV flux). When compared to  RXTE/HEXTE results
(columns 2--3), this is no longer valid. Our observations do not
confirm the steep spectra measured by HEXTE in some occasions.
Also, our fluxes are much lower than the ones reported by HEXTE.

Thanks to our long-term INTEGRAL/IBIS database on Sco{\thinspace}X-1 we were
able to track the behavior of its hard X-ray emission 
($E \; > \; 50${\thinspace}keV).  As shown in Figure{\thinspace}({\ref{fig2}}) the hard X-ray
emission from $50$ to $200${\thinspace}keV has a tight correlation with the
emission between $20$ and $50${\thinspace}keV. Up to this energy the spectral
modeling of our data (as well of \citeads{2006ApJ...649L..91D}) 
is the Comptonization of seed photons of lower energies by a
hot plasma. Due to the compelling correlation of
Figure{\thinspace}({\ref{fig2}}) we thus suggest that the physical
mechanism responsible for the emission up to 200{\thinspace}keV in
Sco{\thinspace}X-1 is Comptonization.

\section{Discussion}
\label{discussion}

\begin{figure}
\resizebox{\hsize}{!}{\includegraphics[angle=270]{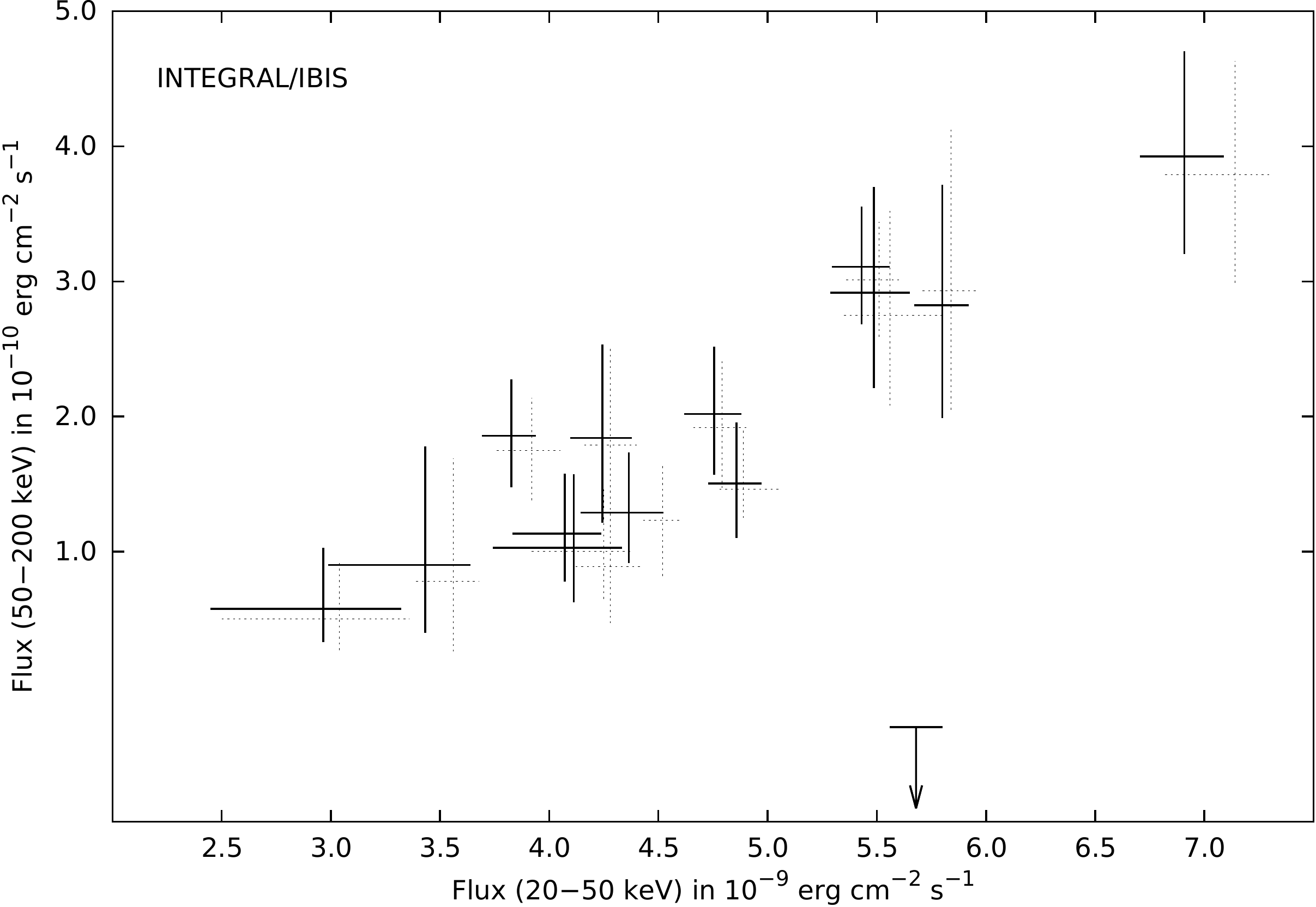}}
	\caption{
                 The powerlaw (solid lines) and {\tt comptt} (dashed lines) 
                 50--200{\thinspace}keV 
                 components plotted against the thermal component (20--50{\thinspace}keV)using the 
                 results displayed in Table ({\ref{tab1}}). As in Figure ({\ref{fig1}}) the arrow 
                 denotes a 3$\sigma$ upper limit 
                 of $\sim \; 5 \; \times \; 10^{-12}$ erg{\thinspace}cm$^{-2}${\thinspace}s$^{-1}$ 
                 to the powerlaw flux.
                }
\label{fig2}
\end{figure}

The INTEGRAL observations of Sco X-1 reported here show that a
nonthermal hard component is almost always needed to fit the Sco
X-1 X-ray spectra, contradicting earlier results based on HEXTE/RXTE
data. 

The most important result of this work is the tight correlation
between the thermal and nonthermal fluxes found in the
IBIS data.  This seems to indicate that the same mechanism responsible
for the thermal component is also at work in the case of the
nonthermal powerlaw component. We thus interpret the high-energy component as
a second Comptonization. 
As we have shown here, see Figure{\thinspace}({\ref{fig2}}),
this second component can be fitted by the Comptonization of seed photons
by a $\sim${\thinspace}$60${\thinspace}keV plasma (a second {\tt{comptt}}). 
In this scenario the spectra up to 50{\thinspace}keV can be interpreted as
Comptonization of seed photons coming from the neutron star itself, by a 
$\sim${\thinspace}$2.71${\thinspace}$\pm${\thinspace}0.04{\thinspace}keV plasma (cloud),
as derived from the values shown in Table{\thinspace}({\ref{tab1}}).
As we have highlighted, a pure blackbody
model also provides an adequate fit to our Sco{\thinspace}X-1 spectra up to 50{\thinspace}keV. 
The second Comptonization component (50 to 200{\thinspace}keV)
must be interpreted as originating in an outer part of the system, requiring
(as expected) an (electron) temperature on the order of $60${\thinspace}keV.  
But as we have shown here, the tight correlation between the fluxes of the two
components is also held with the 50 to 200{\thinspace}keV spectra modeled as a simple
powerlaw (see Figure{\thinspace}{\ref{fig2}}). Based on this correlation, it is our suggestion
that the flux of Sco{\thinspace}X-1 up to 200{\thinspace}keV is due to Comptonization.
Beacause the fits from 50 to 200{\thinspace}keV with a 
{\tt{powerlaw}} model are slightly better (as we have shown, see Table{\thinspace}{\ref{tab1}}),
we are also suggesting that the Comptonization in this part of the spectrum is nonthermal
in origin. 

\begin{figure}
\resizebox{\hsize}{!}{\includegraphics[angle=0]{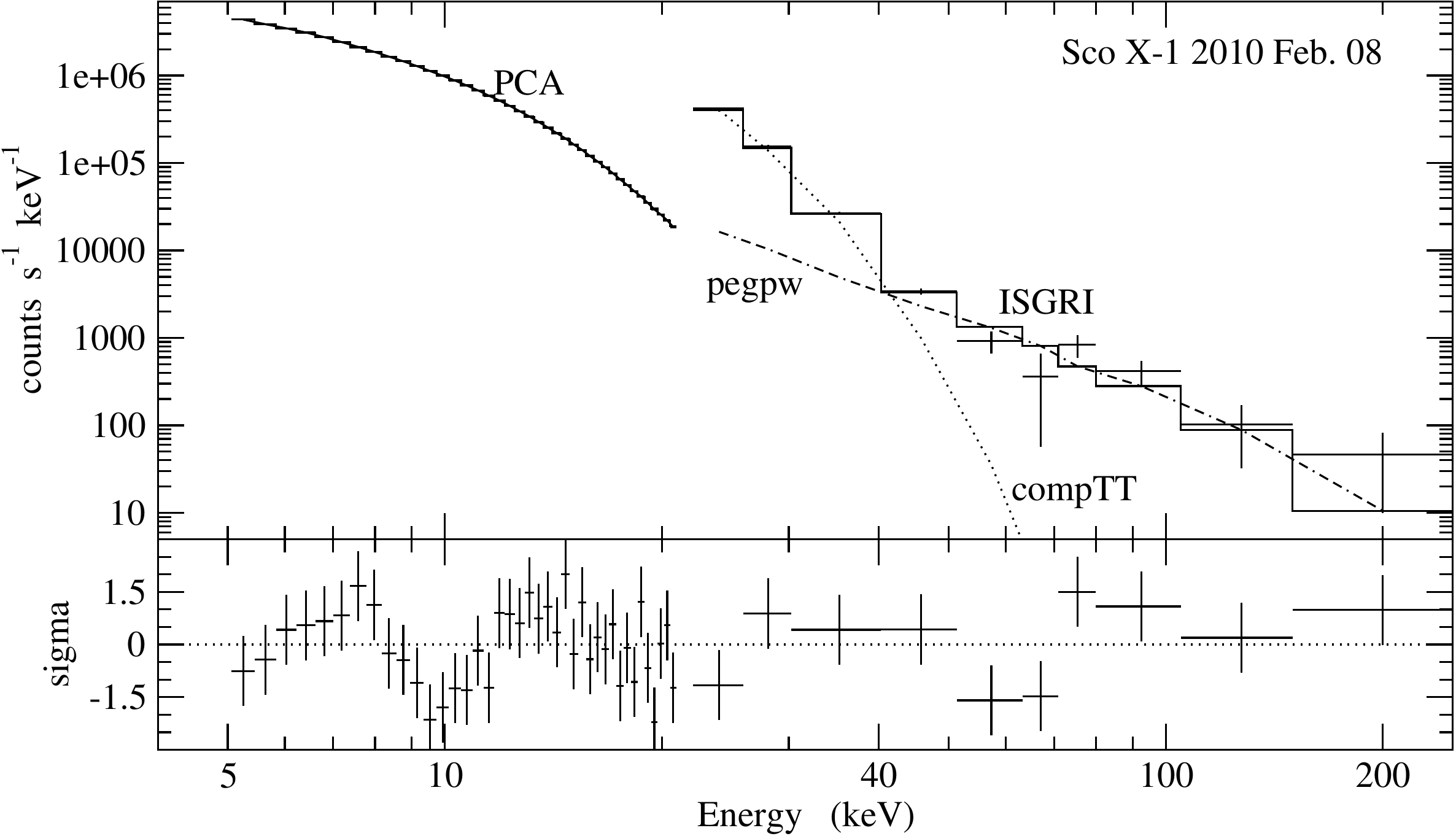}}
	  \caption{
                   A joint PCA/RXTE and IBIS/INTEGRAL observation of Sco{\thinspace}X-1. The spectrum
                   was fitted by a model with a normalization constant (between PCA and 
                   IBIS), photoelectric absorption, plus a comptt and powerlaw components. 
                   Because the PCA cannot constrain the equivalent hydrogen column for the
                   absorption, its value was fixed at a value of 4. The plasma 
                   temperature (derived from comptt model) is 
                   $2.90^{\hbox{\tiny{+0.02}}}_{\hbox{\tiny{-0.01}}}$. The reduced $\chi^2$ is
                   1.3. Also displayed in the figure are the relative contributions (at the IBIS part
                    of the fit) from the {\tt comptt} and {\tt pegpw} components.
        	  }
\label{fig3}
\end{figure}

Figure{\thinspace}({\ref{fig3}}) shows a simultaneous RXTE and INTEGRAL
observation.  The data from $5$ to 50{\thinspace}keV is well fitted by a
single {\tt comptt} component with a temperature of
$2.90^{+0.02}_{-0.01}$. This strengthens our interpretation for the
Comptonization origin of the soft Sco{\thinspace}X-1 spectra. As already 
mentioned, the average plasma temperature derived from the results of
Table{\thinspace}({\ref{tab1}}) is $2.71 \pm 0.04${\thinspace}keV, which is in very
reasonable agreement with the joint fit.

In our database the average powerlaw index is $3.06 \pm 0.21$,
in agreement with values for LMXBs (e.g., \citeads{2006ApJ...649L..91D}). This
result differs, however, from the reported values derived from
long-term HEXTE/RXTE observations (\citeads{2001AdSpR..28..389D}). It is not
clear what causes this discrepancy, but it may be the long-term
intrinsic variability of the source.

Our thermal luminosities are within 
$2.8$--$6.5${\thinspace}$\times${\thinspace}$10^{36}${\thinspace}erg{\thinspace}s$^{-1}$. 
It has been claimed before (\citeads{2001AIPC..587...44D}) 
that the appearance of the powerlaw component in 
{\sf Z} sources is tied to a thermal luminosity greater than 
$4${\thinspace}$\times${\thinspace}$10^{36}${\thinspace}erg{\thinspace}s$^{-1}$ . Our results here, 
as well as others (e.g., \citeads{2006ApJ...649L..91D}), contradict this statement.


\section{Conclusions}
\label{conclusion}

We shown evidence here of  Comptonization up to 200{\thinspace}keV in
Sco{\thinspace}X-1 spectra, which is supported by a strong correlation
between the fluxes from $20$ to $50${\thinspace}keV and 
$50$ to $200${\thinspace}keV. This
(second) Comptonization might be non-thermal in origin.
We have analyzed a seven-year database of IBIS/INTEGRAL data,
which turned out to be the longest IBIS database on Sco{\thinspace}X-1
(besides the 4th source catalog). We found no evidence
of any preferred luminosity state for the thermal component
regarding the appearance of the hard X-ray tail. Our powerlaw indexes
are fully compatible with the reported values for LMXBs. We found no
evidence of any states with ${\Gamma} \; \sim \; 0$ reported before in the
literature. By analyzing a simultaneous PCA/RXTE and IBIS/INTEGRAL
observation of Sco{\thinspace}X-1, we were able to fit the thermal part of the
spectrum from $5$ to $50${\thinspace}keV with a single Comptonization
model. The production of hard X-ray tails in {\sf Z} sources is still
a hot topic of debate, and the power and unique capabilities
of IBIS/INTEGRAL is a fundamental tool for better constraining physical
models of their origins.


\begin{acknowledgements}
TM gratefully acknowledges CAPES-Brazil for financial support. 
We thank an anonymous referee for comments that helped us in
improving the quality of our study.
\end{acknowledgements}

\bibliographystyle{aa}    
\bibliography{tmfdjb}     

\end{document}